\documentclass[12pt]{article}

\usepackage{scicite}
\usepackage{amsmath}
\usepackage{epsfig}
\usepackage{graphicx}
\usepackage{url}

\usepackage{times}

\newenvironment{sciabstract}{%
\begin{quote} \bf}
{\end{quote}}

\newcounter{lastnote}
\newenvironment{scilastnote}{%
\setcounter{lastnote}{\value{enumiv}}%
\addtocounter{lastnote}{+1}%
\begin{list}%
{\arabic{lastnote}.}
{\setlength{\leftmargin}{.22in}}
{\setlength{\labelsep}{.5em}}}
{\end{list}}

\newcommand{\msun}{{\rm M}_{\odot}}     
\newcommand{\mlow}{10^6\,\msun}
\newcommand{\mhi}{10^{11}\,\msun}

\newcommand{\dint}{\displaystyle\int}

\newcommand{\beq}{\begin{equation}}
\newcommand{\eeq}{\end{equation}}
\newcommand{\bea}{\begin{eqnarray}}
\newcommand{\eea}{\end{eqnarray}}

\def\leq{\,\raise 0.4ex\hbox{$<$}\kern -0.8em\lower 0.62ex\hbox{$-$}\,}
\def\geq{\,\raise 0.4ex\hbox{$>$}\kern -0.8em\lower 0.62ex\hbox{$-$}\,}
\def\pm{\,\raise 0.4ex\hbox{$+$}\kern -0.8em\lower 0.62ex\hbox{$-$}\,}
\def\lsim{\,\raise 0.4ex\hbox{$<$}\kern -0.8em\lower 0.62ex\hbox{$\sim$}\,}
\def\gsim{\,\raise 0.4ex\hbox{$>$}\kern -0.8em\lower 0.62ex\hbox{$\sim$}\,}
\def\apppropto{\,\raise 0.4ex\hbox{$\propto$}\kern -0.7em\lower 0.62ex\hbox{$\sim$}\,}

\begin{document}


\title{The imminent detection of gravitational waves from massive black-hole binaries with pulsar timing arrays}


\author{Sean T. McWilliams${}^{1\ast}$, Jeremiah P. Ostriker${}^{2}$ and Frans Pretorius${}^{1}$\\
\normalsize{${}^{1}$Department of Physics, Princeton University, Princeton, NJ 08544 USA}\\
\normalsize{${}^{2}$Department of Astrophysical Sciences, Princeton University, Princeton, NJ 08544 USA}\\
\normalsize{${}^{\ast}$To whom correspondence should be addressed; stmcwill@princeton.edu}}


\date{}



\baselineskip24pt


\maketitle


\begin{sciabstract}
Recent observations of massive galaxies indicate that they double in mass and quintuple in size between redshift $z=1$ and the present,
despite undergoing very little star formation, suggesting that galaxy mergers drive the evolution.
Since these galaxies will contain supermassive black holes, this suggests
a larger black hole merger rate, and therefore a larger gravitational-wave signal, than previously expected.
We calculate the merger-driven evolution of the mass function, and find that merger rates
are $10$--$30$ times higher and gravitational waves are $3$--$5$ times stronger than previously estimated,
so that the gravitational-wave signal may already be detectable with existing data from pulsar timing arrays.
We also provide an explanation for the disagreement with past estimates that were based on dark matter halo simulations.

\end{sciabstract}


Until recently, the massive galaxies living near the centers of clusters were thought to
be ``red and dead'' during the interval $z\leq1$ (equivalently, for $D_L \leq 6.7$ Gpc or the last 7.8 billion years, assuming WMAP's
seven-year cosmological parameters \cite{WMAP7}), meaning they no longer underwent much star formation, and they remained essentially static.
However, a growing body of evidence, both observational \cite{Robaina,vanDokkum,Trujillo} and computational \cite{Naab,Oser}, 
indicates that while this epoch indeed saw little
star formation in massive galaxies, these galaxies were far from static: to the contrary, massive galaxies are seen to have typically doubled their
mass, and quintupled their size during this period \cite{vanDokkum}.
The most likely explanation for the observations is that massive galaxies merge with less massive satellite galaxies, and this is
supported by simulations.  Since these merging galaxies will generally also contain massive black holes, the larger-than-expected galaxy
merger rate also implies an unexpectedly large black-hole binary merger rate.  This has significant implications for pulsar timing arrays (PTAs)
that are currently searching for these gravitational-wave sources.

We therefore calculate the expected gravitational-wave signal from this entire population.  Whereas the most recent estimates
of this signal rely on dark matter halo simulations (e.~g.~\cite{Millenium}) to
provide a population of merging binaries, we instead develop a simple observationally-normalized 
model for merger-driven galaxy evolution.  We expect our resulting
signal estimate to be significantly larger for the following reason: if we treat the satellite galaxy as a stellar bulge inside a halo, rather than
a halo on its own, this has a surprisingly large impact on the mass of the satellite that survives being stripped by the host,
and therefore the dynamical friction timescale $t_{\rm df}$ for the satellite to reach the center of the cluster.  If we use an NFW profile \cite{NFW}
for the halo, it is simple to show that $t_{\rm df} \propto (1+M_{\star}/M_{\rm DM})^{-9}$ \cite{MOP},
where $M_{\star}$ is the stellar mass, and $M_{\rm DM}$ is the mass of the tidally limited subhalo containing the stellar bulge with an
embedded black hole.  Therefore, if after tidal
stripping we find that $M_{\star}/M_{\rm DM}\approx 1/2$, for example, then the satellite will sink $\sim40$ times faster than it would in
a pure dark matter simulation, suggesting such simulations may seriously underestimate merger rates.

\begin{figure}
\begin{center}
\includegraphics[trim = 0mm 0mm 0mm 0mm, clip, scale=.55, angle=0]{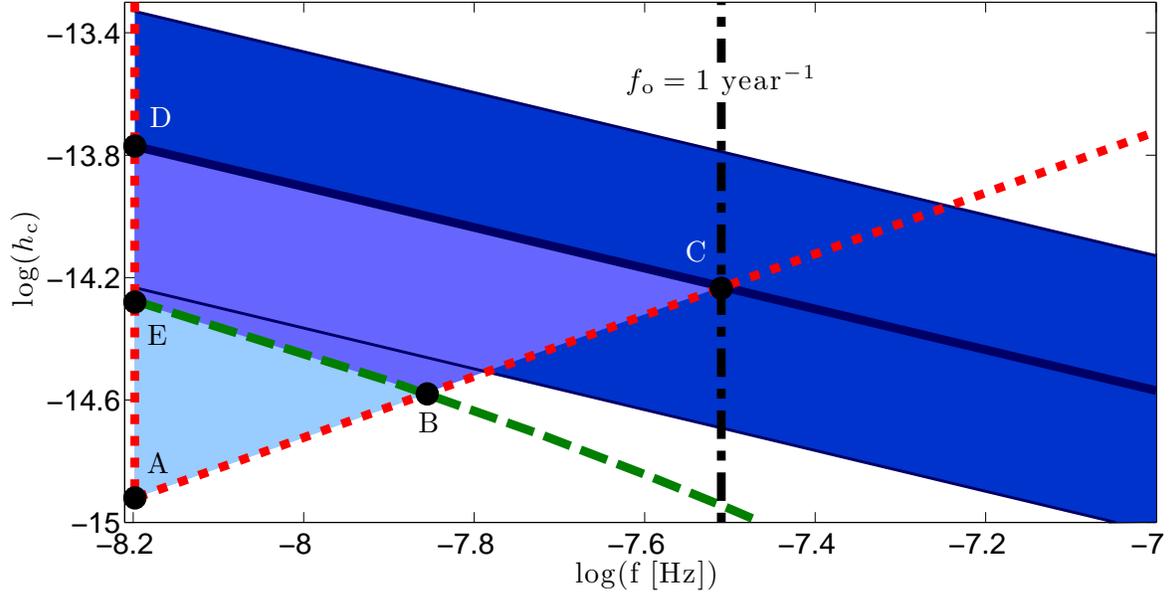}
\caption{Gravitational wave strain and strain sensitivity for a 5 year observation with PTAs. 
The red dashed line is the approximate strain sensitivity for current PTAs \cite{EPTA}, and the green dashed line shows the previous estimate
for the stochastic signal strength that is currently in standard use for PTA analyses \cite{Sesana:2008mz,Sesana2010}.
The dark blue solid line corresponds to our mean estimate for the stochastic signal strength, with the blue shaded region bound
by thin solid blue lines showing our 95\% confidence interval for this estimate, based on the observational uncertainties
of our model parameters.  The light blue ($\triangle ACD$) and cyan ($\triangle ABE$) shaded regions show the area corresponding
to the square root of the $\rho^2$ integrand, to be integrated over logarithmic frequency intervals as in Eq.~\eqref{eq:snr},
for our expected SNR of 8, and the SNR of 2 expected from previous estimates \cite{Sesana:2008mz,Sesana2010}, respectively.}
\label{fig:strain}
\end{center}
\end{figure}

We begin describing our model by noting that under the assumption of negligible star formation, the
probability of a merger between black holes of mass $M'$ and $M''$ during the redshift interval $z$--$z+dz$ is given by
\beq
\frac{\partial^3 p}{\partial M' \partial M'' \partial z} dM' dM'' dz = P\, dz\, \phi(M',z)\, dM' \,\phi(M'',z) \,dM''\,,
\label{eq:prob}
\eeq
where $\phi$ is the black-hole mass function used in \cite{Lin} and $P$ is a normalization constant.  We also note that
the probability of finding a binary in a given frequency
interval $df$ is found by observing that the ``chirp rate'' $\dot{f}\propto f^{11/3}$, so that $\frac{dp}{df}\propto f^{-11/3}$.
We can therefore express the expected signal strength from the full population of binaries with respect to the probability distribution as
\beq
h_{\rm c}^2(f)=\dint_{0}^{1}dz \dint_{\mlow}^{\mhi}dM_2 \dint_{\mlow}^{M_2} dM_1 Nh^2 \frac{d^4p}{dM_1dM_2dz\,df}\,,
\label{eq:hc}
\eeq
where $N$ is the total number of sources, and $h$ is the strain from an individual source.
We implement Eq.~\eqref{eq:hc} through Monte Carlo simulations, summing the power of each selected binary to arrive at the
full stochastic signal power, $h_{\rm c,\,signal}^2$.  The instrumental error from the radiometer is presently 
the limiting noise source for pulsar timing arrays,
and places a limit on the amount of strain, $h_{\rm c,\,noise}$, that is detectable in the data.  The signal-to-noise ratio (SNR), $\rho$,
is simply given by
\beq
\rho^2=\int_0^{\infty} \frac{df}{f} \left(\frac{h_{\rm c,\,signal}}{h_{\rm c,\,noise}}\right)^2\,,
\label{eq:snr}
\eeq
where the strain signal is averaged over
the orientation angles of the binaries and the sensitivity is averaged over the sky position angles.
In Fig.~\ref{fig:strain}, we show the signal strain for our present model and for the current standard result used in PTA analyses,
as well as the noise level for the current generation of PTAs assuming a 5 year observation.  We find that our mean SNR is $\sim 4$ times
the previous estimate; the triangles $\triangle ACD$ and $\triangle ABE$ show the square 
root of the $\rho^2$ integrand, so their areas are visual indicators
of the relative SNRs of our signal and the previous estimate, respectively. 
Our mean signal estimate is $\sim 5 \times$ stronger than the previous estimate at $f_{\rm o}=1\,{\rm year}^{-1}$;
indeed, our mean estimate of $h_{\rm o} \geq 5.8\times 10^{-15}$, where $h_{\rm c}=h_{\rm o}\left(f/1\,{\rm year}^{-1}\right)^{-2/3}$ defines $h_{\rm o}$,
nearly equals the 95\% null confidence limit of
the European PTA ($6\times 10^{-15}$) \cite{EPTA} and the NANOGrav PTA ($7\times 10^{-15}$) \cite{nanograv}. 
This indicates that some portion of the space
of possible observational parameters is already ruled out within our model, and that we expect, in the statistical sense,
that a positive detection will be made with PTAs in the very near future.  If we assume that radiometer noise continues to dominate
over red timing noise, then the noise floor of PTAs will improve with observation time $T$ as $T^{-13/6}$ \cite{nanograv}.  Therefore, by 2016,
PTAs will be sensitive to our most pessimistic set of model parameters which yield $h_{\rm o}=2\times 10^{-15}$, so that we expect
a detection by 2016 with 95\% confidence based on our merger-driven model, with improved data quality and the possible combination of data from different
PTAs only serving to accelerate the discovery.  Since Advanced LIGO will not begin science operation until 2015, and will not reach
its design sensitivity until 2018-2019, it is therefore likely that PTAs will make the first direct detection
of gravitational waves, a landmark discovery which will usher in a new era of observational gravitational-wave astronomy.


\bibliographystyle{Science}
\bibliography{/Users/sean/publish/bibtex/references}

\begin{thebibliography}{10}

\bibitem{WMAP7}
E.~{Komatsu}, {\it et~al.\/}, {\it Astrophys. J. Supp. Ser.\/} {\bf 192}, 18
  (2011).

\bibitem{Robaina}
A.~R. {Robaina}, {\it et~al.\/}, {\it Astrophys. J.\/} {\bf 719}, 844 (2010).

\bibitem{vanDokkum}
P.~G. {van Dokkum}, {\it et~al.\/}, {\it Astrophys. J.\/} {\bf 709}, 1018
  (2010).

\bibitem{Trujillo}
I.~{Trujillo}, I.~{Ferreras}, I.~G. {de La Rosa}, {\it Mon. Not. R. Astron.
  Soc.\/} {\bf 415}, 3903 (2011).

\bibitem{Naab}
T.~{Naab}, P.~H. {Johansson}, J.~P. {Ostriker} {\bf 699}, L178 (2009).

\bibitem{Oser}
L.~{Oser}, J.~P. {Ostriker}, T.~{Naab}, P.~H. {Johansson}, A.~{Burkert}, {\it
  Astrophys. J.\/} {\bf 725}, 2312 (2010).

\bibitem{Millenium}
V.~{Springel}, {\it et~al.\/}, {\it Nature\/} {\bf 435}, 629 (2005).

\bibitem{NFW}
J.~F. {Navarro}, C.~S. {Frenk}, S.~D.~M. {White}, {\it Astrophys. J.\/} {\bf
  462}, 563 (1996).

\bibitem{MOP}
S.~T. McWilliams, J.~P. Ostriker, F.~Pretorius, Gravitational waves and stalled
  satellites from massive galaxy mergers at $z\leq 1$ (2012). In preparation.

\bibitem{Lin}
Y.-T. {Lin}, J.~P. {Ostriker}, C.~J. {Miller}, {\it Astrophys. J.\/} {\bf 715},
  1486 (2010).

\bibitem{EPTA}
R.~{van Haasteren}, {\it et~al.\/}, {\it Mon. Not. R. Astron. Soc.\/} {\bf
  414}, 3117 (2011).

\bibitem{nanograv}
P.~B. {Demorest}, {\it et~al.\/}  (2012). {arXiv:1201.6641 [astro-ph.CO]}.

\bibitem{Sesana:2008mz}
A.~Sesana, A.~Vecchio, C.~N. Colacino, {\it Mon. Not. R. Astron. Soc.\/} {\bf
  390}, 192 (2008).

\bibitem{Sesana2010}
A.~Sesana, A.~Vecchio, {\it Class. Quantum Grav.\/} {\bf 27}, 084016 (2010).

\end{thebibliography}


\begin{scilastnote}
\item We wish to thank Alberto Sesana and Priya Natarajan for helpful discussions.
Correspondence and requests for materials should be addressed to S.T.M. at stmcwill@princeton.edu
\end{scilastnote}



\end{document}